\documentstyle[preprint,eqsecnum,aps]{revtex}
\begin{document}
\draft
\def \beq{\begin{equation}}
\def \eeq{\end{equation}}
\def \beqarr{\begin{eqnarray}}
\def \eeqarr{\end{eqnarray}}

\title{Generalised Chern-Simons Theory of Composite Fermions \\
 in Bilayer Hall Systems}
\author{R. Rajaraman\cite{byline1}}

\address{School of Physical Sciences \\
Jawaharlal Nehru University\\ New Delhi 110067, \ INDIA\\ }

\date{\today}
\maketitle
\begin{abstract}
We present a field theory of Jain's composite fermion model\cite {Jain}, as
generalised to the bilayer quantum Hall systems. We define
operators which create composite fermions and write the
Hamiltonian exactly in terms of these operators. This is seen to
be a complexified version of the familiar Chern Simons theory.
In the mean-field approximation, the composite fermions feel a
modified effective magnetic field exactly as happens in usual
Chern Simons theories, and plateaus are predicted at the same
values of filling factors as Lopez and Fradkin \cite{Lopez3}and
Halperin \cite{Halp}.  But unlike normal Chern Simons theories,
 we obtain all features of the
 first-quantised wavefunctions including its phase, modulus and
 correct gaussian factors at the mean field level. The
familiar Jain relations for monolayers and the Halperin
wavefunction for bilayers come out as special cases.
\end{abstract}
\section{Introduction}

With the development of techniques for growing GaAs
heterostructures containing two separated layers of
two-dimensional electron gas, experimental work on the quantum
Hall effect has been extended to such bilayer systems as well
(see for example references \cite{Eisen} ,\cite {Murphy},
\cite{Suen}, \cite{Boeb}).  New plateaus in Hall conductivity
have been observed at filling fractions not seen in single
layers such as at $\nu = {1\over 2}$. On the theoretical front,
a large body of work has already been done on bilayer systems.
An extensive list of references to this literature has been
given in the lucid review of this subject by Girvin and
MacDonald \cite{GirvMac} and in the paper by Moon et al
\cite{Moon}. In particular, as background and motivation for our
present work let us recapitulate  the following theoretical
developments :
 
 (i) A major step which helped in the study of  quantum Hall
effect was the proposal of elegant and yet very accurate
first-quantised N-particle wavefunctions for the   ground state
and quasiparticle excitations  at the Hall
conductivtity plateaus.  This was pioneered by Laughlin for the mono layer
case with his famous wavefunction for the filling fractions of
$\nu = {1 \over 2m+1}$ \cite{Laugh}. For  double layer systems (viewed as
a two component system carrying a pseudo-spin layer index ) a
generalisation of the Laughlin wavefunction was proposed long
ago by Halperin \cite{Halp} . The Halperin wavefunction
$\psi_{m_{1}, m_{2}, n}$ is labelled by three integers $m_{1},
m_{2}$ and $n$ -- of which $m_1$ and $m_2$ must be odd -- which
determine the filling fraction $\nu$. For example, it was
proposed by Yashioka, McDonald and Girvin \cite{YMG}  that the
 plateau at $\nu = {1 \over 2}$ seen in the
bilayer system corresponded to the wavefunction $\psi_{3,3,1}$.

  (ii) For  mono-layers , the  fractional quantum Hall effect
plateaus and their phenomenologically very successful
wavefunctions  were derived or justified from underlying
composite particle formation postulates.  Jain
\cite{Jain} presented a theory of flux-electron composite
fermion formation. Jain's theory related Hall plateaus and their  electron
wavefunctions at  fractional fillings to corrresponding plateaus
and wavefunctions of the   composite fermions at integral
fillings. 

 (iii)	On a different front, the fractional effect was
studied in field theoretic formulations. Based on the
observations of Girvin and MacDonald \cite {GM} that these
systems seem to exhibit off-diagonal long range order, Zhang,
Hansson and Kivelson \cite{ZHK} constructed a Chern Simons field
theory of a Landau Ginsberg order parameter field for quantum
Hall effect at fractional fillings $\nu = {1
\over 2m+1}$. Their bosonic order parameter
field    corresponds to composites of electrons with
an  odd number of fluxons. A similar Chern Simons field theory, but with even
-integer coupling, was studied by Lopez and Fradkin  \cite{Lopez1}. This
 corresponds to having composites of even number of fluxons with electrons
 and gives a  field theoretic formulation of Jain's theory. Subsequently, 
Lopez and Fradkin \cite{Lopez3} also extended their fermionic Chern Simons 
field theories to the bilayer case and predicted possible Hall plateaus for a
 large family of filling fractions in each layer. For related 
 work on partially polarised electrons  see Mandal and Ravishankar
  \cite{Ravi}. Ezawa and Iwasaki \cite{Ezawa} also studied
bilayer systems through a bosonic Chern Simons theory. They
solved it using self-duality equations which hold in the absence
of Coulomb interactions, which they treat perturbatively in the
short distance limit.
  
 (iii) All these papers in references \cite{ZHK} , \cite{Ezawa},
 \cite{Lopez1} and \cite{Lopez3} 
 extracted first quantised N-particle wavefunctions from their field
theoretic ground states.  But in the mean field approximation,
only some aspects of the Laughlin wavefunction emerged. In the
Zhang et al theory \cite{ZHK}, the Landau Ginsberg field
incorporates only the correct phases of the electronic
correlations and not their modulus. The all important zeroes in
the Jastrow correlation factors as well as the gaussian factors
in the Laughlin wave functions emerge only upon including
fluctations about the mean field theory. In ref.\cite {Lopez1} the
modulus of the Laughlin wavefunction is derived by a very
different and ingenious method starting from the low-$q^2$ limit
of correlation functions.  But such a result holds only in the
long distance limit.  The same is also true of the Lopez
-Fradkin bilayer work \cite{Lopez3} where again they have obtained the
Halperin-Jain wavefunction for some cases, but only in the long
distance limit. The wavefunctions obtained in the Ezawa-Iwasaki work
hold only at short distances and are in the absence of Coulomb interactions.

In this paper we present a modified  Chern Simons field theory
for bilayer systems, which yields naturally at the mean field
level Jain's model of composite fermion formation as generalised
to bilayer systems as well as the corresponding wavefunctions.
 (Jain's well known monolayer results also come
out as a special case.)
 The present work is an adaptation to bilayer fermionic
composite operators of our earlier work with Sondhi \cite{Raj}
where we had  presented an exact field theory of the Read
operator \cite {Read}. There we had employed a  complexified version of the
 Chern Simons Landau Ginsberg field theory 
 which enabled us to reproduce all the features of the Laughlin
wavefunction already at the mean field level. We use an appropriate 
generalisation of the same  method here.
 We  explicitly construct operators that
create bilayer composite fermions, using a non-unitary
transformation acting on the parent electron field operator .
Exact anti-commutation rules and an exact Hamiltonian are
written in terms of these composite fermion operators. In the
mean field approximation this Hamiltonian relates electrons at
fractional fillings to composite fermions at integer fillings.
These fractions and integers are seen to be related precisely by the
formulae given by Lopez and Fradkin.  Equations are also
obtained akin to Jain's , but generalised to bilayers, which
relate electron wavefunctions to composite fermion
wavefunctions.  The Halperin  wavefunctions are obtained as
specific examples. 

 In our theory, all aspects of these wavefunctions -- the
phases and moduli of their Jastrow correlations, and  appropriate
gaussian factors -- emerge in the lowest order of the mean-field
approximation. No short distance or long distance approximation is used nor any
lowest Landau level restriction put in by hand. That our method of reference
 \cite{Raj} can be used to generate composite fermions was also pointed out by 
 Wu and Yu \cite{Wu} for the monolayer case.

\section{Composite Fermion Operators}

Consider a double layer of two-dimensional electrons of mass 
$\mu$ and charge $e$, 
placed in a uniform perpendicular magnetic field of strength $B$
which corresponds to a vector potential  in the symmetric gauge
of $ {\vec A}({\vec x}) = {1 \over 2 } B  \ {\hat k} \times {\vec x}$, where 
${\hat k}$ is the unit vector perpendicular to the plane.
We take the electrons to be fully spin polarised along the $B$ field
 and hence suppress  spin  for simplicity. Suppose
 their interaction potential is $V({\vec x} - {\vec x'})$ and  the
scalar potential $A_0$  represents any uniform and/or random impurity
electric fields in the problem. Let $ \Psi_{\alpha} ({\vec x})$
denote the bilayer electron quantum field with $\alpha \ = \ 1,2$ 
standing for the layer index. It obeys 
 the equal-time anticommutation relations,
\beqarr 
\{ \Psi_{\alpha}({\vec x}) , \Psi_{\beta}({\vec x}') \} 
 &=& \{ \Psi_{\alpha}^{\dagger} ({\vec x}) ,
 \Psi^{\dagger}_{\beta}({\vec x}') \}  =  0 \nonumber \\ 
\{\Psi_{\alpha}({\vec x}),\Psi^{\dagger}_{\beta}({\vec x}') \} &=& 
\delta_{\alpha, \beta} \ \delta^{(2)} (\vec {x}-\vec{x}') \ .
\eeqarr
Clearly $\Psi_{\alpha}$ is the full electron field and not just its
lowest Landau level projected part. 
Note that unless explicitly specified, it is understood throughout this 
article that repeated indices $\alpha, \beta$ etc. are
 $\underline{not}$ to be summed over.
The  second quantized
Hamiltonian that describes our system is,
\beq 
H  =  \int d^{2}x\, \sum_{\alpha} \bigg[\Psi^{\dagger}_{\alpha}
({\vec x})\, \bigg({-\hbar ^{2} \over 2\mu} {\vec D}^{2}   + eA_0 \bigg)\, 
 \Psi_{\alpha} ({\vec x})  \ \bigg]    +  \ \sum_{\alpha, \beta}  \ 
  {1 \over 2} \int \int d^2 x\, d^2 x'  \, 
\delta\rho_{\alpha}({\vec x})\,  V_{\alpha \beta}(\vec {x}-\vec {x}') \,
 \delta \rho_{\beta}({\vec x}') \ . \label{Ham} 
\eeq  
Here, $ {\vec D} \equiv {\vec \nabla}  -  i 
{e \over {\hbar c}} {\vec A}$ is the covariant derivative inclusive of the 
vector potential of the uniform magnetic field B .
 $\rho_{\alpha} ({\vec x}) \ \equiv  \ \Psi_{\alpha}^{\dagger}({\vec x}) 
\Psi_{\alpha}({\vec x})$ 
is the electron density operator in the layer $\alpha$
 whose deviation from its mean value  \ $\bar 
\rho_{\alpha}$ is $ \delta \rho_{\alpha} ({\vec x})$.

Next we define our operators for the composite fermion field $\chi
_{\alpha}({\vec x})$
and its canonical conjugate field $\Pi_{\alpha}({\vec x})$, by
\beqarr 
\chi_{\alpha}({\vec x}) &\equiv&   e^{ -J
_{\alpha}({\vec x})} \Psi_{\alpha}({\vec x})
 \nonumber \\
\Pi_{\alpha}({\vec x}) &\equiv& \Psi_{\alpha}^{\dagger}({\vec x})
 e^{ J_{\alpha}({\vec x})}, 
\label{chipi} 
\eeqarr
where,
\beq J_{\alpha}({\vec x}) \equiv  \sum_{\beta} R_{\alpha \beta} \int d^{2}x'\,
[\rho_{ \beta}({\vec x}')  \log(z-z')]
   - \   r_{\alpha}{\mid z \mid ^2 \over 4l^2}  , \label{defJ} 
\eeq
In the above equation $r_{\alpha}$ and the $2 \times 2$ matrix
 $R_{\alpha \beta}$ consist of
some numbers which will be fixed shortly, \ 
  $z \equiv x_1+ix_2$  is the complex
coordinate on the plane and $l  =  \sqrt{{\hbar c \over eB} }$ is the
magnetic length. Notice that $\chi_{\alpha}$ and $\Pi_{\alpha}$
 are not hermitian conjugates of each other since
 $J_{\alpha}$ has both hermitian and anti-hermitian pieces.
Nevertheless, as we now show, they form a pair of  canonically conjugate 
Fermi fields, one in each layer.

 Note that the only operator appearing in $J_{\alpha}(\vec x)$ 
is the electron density $\rho_{\beta} ({\vec x}) =
\Psi_{\beta}^{\dagger}({\vec x}) \Psi_{\beta}({\vec x})$,
which obeys the commutation relation,
\beq 
\big[ \rho_{\alpha}({\vec x}), \Psi_{\beta}({\vec x}')\big]  =  
- \delta_{\alpha \beta} \ \Psi_{\alpha}({\vec x})\, 
\delta^2 ({\vec x}-{\vec x}') \ \label{rhopsi}
\eeq
Therefore the following identities follow :
\beqarr 
e^{-J_{\alpha}({\vec x})}\ \  \Psi_{\beta}({\vec x}') &=& (z-z')^{ 
 R_{\alpha \beta}} \ \Psi_{\beta}({\vec x}') \ 
e^{-J_{\alpha}({\vec x})} \nonumber  \\ 
\Psi_{\beta}^\dagger({\vec x}')\, e^{-J_{\alpha}({\vec x})} &=& (z-z')^{
R_{\alpha \beta}}
 e^{-J_{\alpha}({\vec x})} \, \Psi^\dagger_{\beta}({\vec x}')
\ . \label{ident}
\eeqarr
Using these identities one can verify that
\beq 
\chi_{\alpha}({\vec x}) \chi_{\beta}({\vec x}')  =  (-1)^{R_{\alpha \beta}
 + 1} (z-z')^{R_{\alpha \beta} \ - \ 
 R_{\beta \alpha}} \ \ 
\chi_{\beta}({\vec x'}) \chi_{\alpha}({\vec x}')
\label{comm1} \eeq
We can see that the explicit z-z' dependence in (\ref{comm1}) drops out if
$R_{\alpha \beta}$ is chosen to be a symmetric matrix. Further, if its diagonal
elements are even intergers ($R_{11} = 2s_1$ and $R_{22} = 2s_2$) then
the field     $\chi$ in each layer anticommutes with itself , as desired
of composite fermions.  The off diagonal element $R_{12} = R_{21}$
can be taken to be an integer n. Depending on whether n is 
odd (even), the fields at two different layers will commute (anticommute).  
In short the requirement that the composite fields defined in eq (\ref{chipi})
be fermi fields restricts the matrix $R_{\alpha \beta}$ to have the form
\beq
R_{\alpha \beta} \ = \ \pmatrix{2s_1&n\cr
n&2s_2\cr} \label{matrix}\eeq
exactly in accordance with ref \cite{Lopez3}.

The same choice of $R_{\alpha \beta}$ also yields the canonical anti-commutator
between $\chi_{\alpha}$ and $\Pi_{\beta}$. We have,upon using the identities
(\ref{ident}),
 \beqarr \chi_{\alpha}({\vec x}) \Pi_{\beta}({\vec x}')  &=&
 e^{ -J_{\alpha}({\vec x})} \Psi_{\alpha}({\vec x})
\Psi_{\beta}^{\dagger}({\vec x'})
 e^{ J_{\beta}({\vec x'})} \nonumber \\
 &=&  (-1)^{R_{\alpha \beta}+1} \ \Pi_{\beta}({\vec x}')\chi_{\alpha}({\vec x})
  \ + \ \delta_{\alpha \beta}   \delta^2 ({\vec x}-{\vec x}') \ .
\label{comm2} \eeqarr
Thus, since $R_{11}$ and $R_{22}$ have been even integers, the fields
 $\chi_{\alpha}$ and $\Pi_{\alpha}$ form a pair of 
mutually conjugate Fermi fields for each layer $\alpha$.
This is despite the presence of non-unitary factors in their definition in 
Eq. (\ref{chipi}).
However, in contrast to standard fermi field theories, here
 $\Pi_{\alpha}$ is not equal to $\chi_{\alpha}^{\dagger}$. Instead they
  obey the more complicated relation  
\beq \Pi_{\alpha}({\vec x}) = \chi^\dagger_{\alpha}({\vec x}) 
e^{ J_{\alpha}({\vec x}) + J^\dagger_{\alpha}({\vec x})} \ .
\label{chipirelation}
\eeq
  This has to be borne in mind in doing
manipulations with these composite fermion fields. 

We will define the composite fermion density  $\rho_{\alpha}$ by
\beq 
\rho_{\alpha}({\vec x})  =  
\Pi_{\alpha}({\vec x}) \chi_{\alpha}({\vec x}) \ .  \label{rhocomp} 
\eeq
The corresponding number operator ${\hat N}_{\alpha} \equiv \int d^2x\,
 \rho_{\alpha}$
 satisfies \beq 
\big[ \  {\hat N}_{\alpha} , \   \Pi_{\beta}({\vec x})  \ \big]  = 
 \delta_{\alpha \beta} \Pi_{\alpha}({\vec x}), \label{Npi} \eeq
i.e. the operator $\Pi_{\alpha}({\vec x})$ creats one extra composite fermion
in the ${\alpha}^{th}$ layer. Notice that we have used the same symbol
$\rho_{\alpha}({\vec x})$ for the this composite fermion density 
as we did for the original electron density since the definition
 given in (\ref{rhocomp}) also satisfies
 \beq  \rho_{\alpha}({\vec x}) \ = \ \Psi_{\alpha}^{\dagger}({\vec x}) \ 
\Psi_{\alpha}({\vec x})  \label{equalrhos} \eeq  

It should be emphasized that our composite fermion operators $
\Pi_{\alpha}({\vec x}) $ and $\chi_{\alpha}({\vec x})$ are defined
 over the same  space-time domain as the original electron
field, as is natural in any field operator transformation in a field
theory. This means that the area of the Hall sample is the same,
whether we consider electrons or composite fermions.
Hence, since the densities of both types of fermions has been
shown to be the same, the total number of composite fermions is
also the same as the number of the original fermions.
Note that in obtaining the identities (\ref{ident}) we have used the 
expression (\ref{equalrhos}) for the density in terms of the electron field
and the associated commutator. If we were instead to use the expression 
(\ref{rhocomp}) in terms the composite fermion operators and the associated
commutators, we can analogously obtain the identities

\beqarr 
e^{-J_{\alpha}({\vec x})}\ \  \chi_{\beta}({\vec x}') &=& (z-z')^{ 
 R_{\alpha \beta}} \ \chi_{\beta}({\vec x}') \ 
e^{-J_{\alpha}({\vec x})} \nonumber  \\ 
\Pi_{\beta}({\vec x}')\, e^{-J_{\alpha}({\vec x})} &=& (z-z')^{
R_{\alpha \beta}}
 e^{-J_{\alpha}({\vec x})} \, \Pi_{\beta}({\vec x}')
\ . \label{ident2}
\eeqarr

Having defined the composite fermion fields and obtained their 
commutation relations, let us next rewrite the Hamiltonian (\ref{Ham}) in
terms of them. First consider the covariant derivative on the electron field.
We have,
\beqarr  {\vec D} \Psi_{\alpha}(x) \ &=& \ {\vec D} \big( e^{J_{\alpha}
({\vec x})} \chi_{\alpha}
 ({\vec x}) \big) \nonumber \\
 \ &=& \ \big({\vec \nabla} \ - \ i {e \over {\hbar c}} {\vec A}({\vec x})\big)
  \  \big( e^{J_{\alpha}({\vec x})} \chi_{\alpha} ({\vec x}) \big) \nonumber \\
  &=& \ e^{J_{\alpha}({\vec x})} \ \ \big( {\vec \nabla} \ - \ 
  i {e \over {\hbar c}}
   {\vec A}({\vec x}) \ + \ {\vec \nabla} J_{\alpha}({\vec x}) \big )
    \ \chi_{\alpha}({\vec x}) \nonumber \\
 \   &=&  \ e^{J_{\alpha}({\vec x})} \big ( {\vec D} -{ie \over {\hbar c}}
{\vec v_{\alpha}}({\vec x})   \big) \  \chi_{\alpha}({\vec x}) \eeqarr
where,
\beq 
{\vec v_{\alpha}}({\vec x}) \equiv  i{\hbar c \over e}
{\vec \nabla}J_{\alpha}({\vec x})  \ .
\label{defa}
\eeq
Hence,
\beq 
D^2 \Psi_{\alpha}  =  e^{J_{\alpha}} \ \big( {\vec D}  -  i{e \over \hbar c}
 {\vec v_{\alpha}} \big)^2  \chi_{\alpha} \eeq
Inserting this into the starting Hamiltonian (\ref{Ham} ), and using
Eqs. (\ref{chipi}) and (\ref{rhocomp}) we get,
\beqarr 
H  &=& \int  d^2x  \, \bigg[\Pi_{\alpha} ({\vec x}) \,
\bigg( {-\hbar ^{2} \over 2\mu}   \big({\vec \nabla}  -
 {ie \over \hbar c} \ (\vec {A} + \vec {v_{\alpha}})
\big)^{2}  +  eA_0 \bigg) \chi_{\alpha} ({\vec x}) \bigg] \ \nonumber \\
& +&  {1 \over 2}  \int  \int \, d^2x \, d^2 x'
\,\delta\rho_{\alpha}({\vec x})\, \ V_{\alpha \beta}({\vec x}-{\vec x}') \,
 \delta\rho_{\beta} ({\vec x}')
\label{HB}   
\eeqarr
This Hamiltonian in terms of the composite fermion fields defined
in (\ref{chipi})  is exactly equal to that of our original electron
problem . No approximations have been made so far. Clearly this is very
similar to a Chern Simons theory but it is more than just a direct 
generalisation to bilayers. As in normal Chern Simons theories,
  the vector field ${\vec v}_{\alpha}$ appearing in (\ref{HB}) above is
also constrained in terms of the density by Eq. (\ref{defa}),
where $J_{\alpha}({\vec x})$ is defined in (\ref{defJ}). But since this
 $J_{\alpha}({\vec x})$
involves more than just the phase of $(z-z')$, the field ${\vec v}$ is not
the bilayer statistical Chern-Simon gauge field  used, for instance,
in \cite{Lopez3}. Because $J_{\alpha}({\vec x})$ has real parts, 
$ {\vec v }_{\alpha}$
is a complex vector field. However,
${\vec v}_{\alpha}$ will turn out to be simply related to Chern Simons fields.

Let us define  a Chern-Simons field for each layer index
$\alpha$ by
\beq {\vec a}_{\alpha}({\vec x})   =  {-\hbar c \over e} {\vec \nabla}_x 
\int \, d^2 x' \, \rho_{\alpha}({\vec x}') \, {\rm Im}  \log (z-z')   \ ,
\eeq  or equivalently
\beq b_{\alpha}({\vec x}) \equiv  \nabla \times  \ 
{\vec a}_{\alpha}({\vec x}) \ = \ - \ \phi_0 \rho_{\alpha}({\vec x}) 
\label{defacs} \eeq
where $\phi_0 \equiv {hc \over e }$ is the flux quantum.
Following reference \cite {Raj},  \ use the  Cauchy-Riemann condition
\beq {\vec \nabla} ( {\rm Re} \log z)  =  {\vec \nabla} ( {\rm Im} \log z)
\times {\hat k} \eeq
where  ${\hat k}$ is a unit vector perpendicular to the plane.
Using this we get,
\beqarr 
{\vec v}_{\alpha}({\vec x})  &=&  {i \hbar c \over e} {\vec \nabla}
 J_{\alpha}({\vec x}) \nonumber \\
&=&  {i \hbar c \over e} \  {\vec \nabla}_x   
  \  \bigg \{ \sum_{\beta} R_{\alpha \beta}  \ \int d^2 x' \,
  \big[\rho_{\beta}({\vec x}')
  ({\rm Re} \log (z-z')  +  i\, 
{\rm Im} \log (z-z'))
\big] \ - \ r_{\alpha} {\mid z \mid ^2 \over {4 l^2}}  \bigg\} \nonumber \\
 &=& \sum_{\beta} R_
 {\alpha \beta} \bigg({\vec a}_{\beta}({\vec x}) 
  +  i \ {\hat k} \times
  {\vec a}_{\beta}({\vec x})  \bigg) -  i \phi_{0} r_{\alpha} 
{{\vec x} \over 4\pi l^2} \ .\label{full v} 
\eeqarr

\section{Mean Field Approximation}

Thus far everything is exact. Let us now introduce the mean field (MF)
approximation, by replacing the actual space dependent density 
operator $\rho_{\beta}({\vec x})$ in (\ref{full v}) by its average value 
$\bar{\rho}_{\beta} = {\nu_{\beta} B \over \phi_0 }$
 where $\nu_{\beta}$ is the filling factor in the layer $\beta$.
 Under this approximation eq.(\ref{defacs}) becomes
\beqarr \big(b_{\alpha}\big)_{MF} &\equiv&  \nabla \times  \ 
\big({\vec a}_{\alpha}\big)_ {MF} \nonumber \\
&=& \ - \ \phi_0 \bar{\rho}_{\alpha} \nonumber \\
&=& - \nu_{\alpha} \ B  \eeqarr 
or equivalently,
\beq \big({\vec a}_{\alpha}\big)_ {MF} \ = \ - \nu_{\alpha} \ {\vec A} \eeq
where ${\vec A}$ is the applied external vector potential. Then, 
(\ref{full v}) reduces to
\beq {\vec v}_{\alpha} \ = \ 
 \ - \ \ {\vec A} \ \sum_{\beta} R_
 {\alpha \beta} {\nu}_{\beta}  
  \  +   \ i \ ({\hat k} \times
  {\vec A})  \bigg( - \sum_{\beta} R_
 {\alpha \beta} {\nu}_{\beta}  \ + \   r_{\alpha} \bigg) \label{vMF} \eeq
 Here we have used the fact that in our symmetric gauge
\beq \phi_0 {{\vec x} \over 4\pi l^2} \ = \ - \ {\hat k} \times {\vec A} \eeq
Then the covariant derivative in the composite fermion Hamiltonian (\ref{HB})
becomes, upon using the MF approximation (\ref{vMF}),
\beqarr {\vec D}_{\alpha} \ &=& \ {\vec \nabla} \  \   - {ie \over \hbar c} 
\ ({\vec A} + {\vec v}_{\alpha}) \nonumber \\
 \ &=& \ {\vec \nabla}  - {ie \over \hbar c} \ {\vec A} \bigg( \ 1 \ - \ 
  \ \sum_{\beta} R_
 {\alpha \beta} \nu_{\beta} \bigg) 
  \  +   \ i \ ({\hat k} \times
  {\vec A})  \bigg( - \sum_{\beta} R_
 {\alpha \beta} \nu_{\beta}  \ + \   r_{\alpha} \bigg) \eeqarr
Recall that the numbers $r_{\alpha}$ \ were introduced in our definition of the
composite field operators (\ref{chipi}) and (\ref{defJ}).  Thus far we had left
them unspecified, but let us now choose them  to satisfy
\beq r_{\alpha} \ = \  \sum_{\beta} R_{\alpha \beta} \nu_{\beta} 
\label{reqn} \eeq
Then the covariant derivative simplifies to
\beq   {\vec D}_{\alpha} \ = \ 
\ \ = \ {\vec \nabla}  - {ie \over \hbar c} \ {\vec A} \bigg( \ 1 \ - \ 
  \ \sum_{\beta} R_
 {\alpha \beta} \nu_{\beta} \bigg) \eeq
Therefore in the MF approximation, our composite fermion field experiences 
a Hamiltonian 
\beqarr H  \ \  = \int  d^2x  \, \bigg[\Pi_{\alpha} ({\vec x}) \,
\bigg( {-\hbar ^{2} \over 2\mu}   \big({\vec \nabla}  -
 {ie \over \hbar c} \ ({\vec {\cal A}}_{\alpha})
\big)^{2}  +  eA_0 \bigg) \chi_{\alpha} ({\vec x}) \bigg] \nonumber \\
+  {1 \over 2}  \int  \int \, d^2x \, d^2 x'
\,\delta\rho_{\alpha}({\vec x})\, \ V_{\alpha \beta}({\vec x}-{\vec x}') \,
 \delta\rho_{\beta} ({\vec x}') \label{H-MF}   \eeqarr
where \beq {\vec{\cal A}}_{\alpha}  \ \equiv \    
\bigg( \ 1 \ - \   \ \sum_{\beta} R_{\alpha \beta} {\nu}_{\beta} \bigg)
 \ \vec A . \eeq
\beq B^{*}_{\alpha} \ \equiv \ curl {\vec{\cal A}}_{\alpha} \
 = \bigg( \ 1 \ - \   \ 
\sum_{\beta} R_{\alpha \beta} {\nu}_{\beta} \bigg) \ B \label{B*} \eeq
is the  effective magnetic field felt by the composite fermions of the
$\alpha$th layer. 
Since the filling factor is inversely proportional to the magnetic field,
this reduced effective magnetic field amounts to a correpondingly
 enhanced filling factor for the composite fermions given by 
\beq p_{\alpha}  \ \equiv \ \bigg( {\nu_{\alpha} \over 1 \ - \ 
  \ \sum_{\beta}  \ R_{\alpha \beta} {\nu}_{\beta}}  \bigg) 
   \label{p-nu} \eeq
When these equations are inverted we get
\beqarr \nu_{1} \ &=& \ {1 \over \Delta} \bigg( n \ - \ {1 \over p_2 } - 2s_2
 \bigg) \nonumber \\
  \nu_{2} \ &=& \ {1 \over \Delta} \bigg( n \ - \ {1 \over p_1 }
 - 2s_1 \bigg) \label{nup-rel}  \eeqarr             
where \beq \Delta \ \equiv \ n^2 \ - \ \bigg(  {1 \over p_1 } + 2s_1 \bigg)
\bigg(  {1 \over p_2 } + 2s_2 \bigg) \eeq
These are precisely the filling factors obtained for the bilayer system
by Lopez and Fradkin \cite{Lopez3} .

\section{First-Quantised Wavefunctions}
	The whole purpose of our defining composite fermion operators is
to give a field theoretic version of Jain's theory of fractional
quantum Hall effect as generalised to bilayer systems.
Following Jain's philosophy, if the effective filling factors
$p_{\alpha}$  are integers then one may intiutively expect the
composite fermions to form incompressible quantum Hall ground
states. In terms of electron coordinates the same quantum Hall
state would appear at fractional filling factor $\nu_{\alpha}$
related to the  $p_{\alpha}$  by equations (\ref{nup-rel}).

Further, since the composite fermion field $\chi_{\alpha}$ can be expressed in
terms of the original electron field $\Psi_{\alpha}$ through equation
(\ref{chipi}), if one knows the first quantised wavefunction of
composite fermions in a given state, the corresponding
wavefunction of electrons in the  same state is directly
obtainable.

Let $\phi_{p_{1},p_{2}} (z_1 , z_2 ,. . . . . . . , z_{N_1} ; \ w_1 , w_2 , .
. . . . .  . , w_{N_2})$ represent the composite fermion
wavefunction at composite fermion filling fractions $p_{\alpha} \ = \ 
{\rho_{\alpha} hc \over eB^{*}_{\alpha}} $, where $B^{*}_{\alpha}$
 is the effective magnetic field felt by the composite fermions in layer 
 $\alpha$ as given in
in eq (\ref{B*}). The $z_i$ are  complex coordinates
 of the composite fermions
on the first layer and the $w_i$ the complex coordinates in the second
layer. Note that $N_1$and $N_2$, which stand for the number of
composite particles in each of the two layers respectively, are
also the number of electrons in each of the two layers, since
the density and Number operators of the electrons and those of
the composite fermions are equal to one another (see eq
\ref{equalrhos}). We recall that both the electron and the composite fermion
operators are defined in the same two-dimensional space with the 
same area (sample size). This first quantised wavefunction can be
written in terms of the field theoretic states and field operators 
$\chi_{\alpha}$ as follows : 
\beqarr \phi_{p_{1},p_{2}} (z_1 , z_2 ,. . . . . . . , z_{N_1} 
&;& \ w_1 , w_2 , . . . . . .  . , w_{N_2})  \nonumber \\
    \ \ &\equiv& \ \ \langle O \vert \chi_{1}(z_1) \chi_{1}(z_2)
 . . . \chi_{1}(z_{N_1}) \ \chi_{2}(w_1) \chi_{2}(w_2) . . . . . .\chi_{2}
 (w_{N_2}) \vert MF \rangle \label{phifn} \eeqarr
where  $\vert MF \rangle$ stands for the mean-field ground state
of the composite fermion system at filling factor $p_{\alpha}$
and $\langle O \vert$ is the vacuum state. Meanwhile corresponding to the
same state $\vert MF \rangle$ the first quantised
wavefunction of the \underline{electrons} (whose filling fractions in the 
two layers are respectively $\nu_{1}, \nu_{2}$) is given by
\beqarr \psi_{\nu_{1},\nu_{2}} (z_1 , z_2 ,. . . . . . . , z_{N_1} &;& \ w_1 , w_2 , . . . . . . 
 . , w_{N_2})  \nonumber \\
  \ &\equiv& \ \ \langle O \vert \Psi_{1}(z_1) \Psi_{1}(z_2)
 . . . \Psi_{1}(z_{N_1}) \ \Psi_{2}(w_1) \Psi_{2}(w_2) . . . . . .\Psi_{2}
 (w_{N_2}) \vert MF \rangle \eeqarr
To relate the two wavefunctions, one only needs to write the operator
$\psi_{\alpha}$ in terms of $\chi_{\alpha}$ using (\ref{chipi}) . We get
\beqarr \psi_{\nu_{1},\nu_{2}} ( z_1 , z_2 ,. . . . . . . , &z& _{N_1} ;
 \ w_1 , w_2 , . . . . . .  . , w_{N_2})  \nonumber \\
 \ = \  \langle O \vert  e^{J_{1}(z_1)} & \chi & _{1}(z_1) e^{J_{1}(z_2)}
   \chi_{1}(z_2)
 . . . e^{J_{1}(z_{N_1})} \chi_{1}(z_{N_1}) \ \nonumber \\
 &e& ^{J_{2}(w_1)} \chi_{2}(w_1) e^{J_{2}(w_2)} \chi_{2}(w_2) . . . . . .
 e^{J_{2}(w_{N_2})} \chi_{2} (w_{N_2}) \vert MF \rangle \eeqarr

Next bring all the $e^J$ factors in the above expression to the
left by commuting them across the operators $\chi$ using the
commutators (\ref{ident2}). We get
\beqarr \psi_{\nu_{1},\nu_{2}} (z_1 , z_2 ,. . . . . . . , z_{N_1}
 &;& \ w_1 , w_2 , . . . . . .  . , w_{N_2})  \nonumber \\
 &=& \prod_{i>j}^{N_1} (z_i - z_j)^{R_{11}} \prod_{k>l}^{N_2}
  (w_k - w_l)^{R_{22}} \prod_{r}^{N_1} \prod_{s}^{N_2} (z_r - w_s)^{R_{12}}
  \nonumber \\
  \ & \langle & O \vert  exp  \big( \sum_{1}^{N_1} 
    J_1 (z_i) \ + \  \sum_{1}^{N_2} J_2 (w_j) \big) \nonumber \\
  &\chi&_{1}(z_1) \chi_{1}(z_2)
 . . . \chi_{1}(z_{N_1}) \ \chi_{2}(w_1) \chi_{2}(w_2) . . . . . .\chi_{2}
 (w_{N_2}) \vert MF \rangle \eeqarr
Next apply the operators $e^J$ to the left on the vacuum state.
Notice from its definition in eq.(\ref{defJ}) that the only nontrivial operator
contained in $J_{\alpha}$ is the density, in the first term. The second
term in J is just a c-number.
In an interacting field theory, the vacuum is not generally an eigenstate
 of the density operator.   However in the spirit of the mean field 
 approximation
 being employed in this section, one can replace the density operator
  by its mean value. The mean density of the
 vacuum is zero. Thus when the operators $e^{J_{\alpha}}$ in the above 
 equation act on the
  left on the vacuum state the density dependent first term of $J_{\alpha}$
  can be taken as zero and only the second ( c-number ) survives , giving
  gaussian factors. Therefore
\beqarr \psi_{\nu_{1},\nu_{2}} (z_1 , z_2 ,. . . . . . . , z_{N_1}
 &;& \ w_1 , w_2 , . . . . . .  . , w_{N_2})  \nonumber \\
 &=& \prod_{i>j}^{N_1} (z_i - z_j)^{2s_{1}} \prod_{k>l}^{N_2}
  (w_k - w_l)^{2s_{2}} \prod_{r}^{N_1} \prod_{s}^{N_2} (z_r - w_s)^{n}
  \nonumber \\
&exp& \bigg[ {-1 \over 4l^2} \big(r_1 \sum_{1}^{N_1} 
    |z_i|^2 \ + \ r_2 \sum_{1}^{N_2} |w_i|^2) \bigg]\nonumber \\
  \ & \langle & O \vert  
  \chi_{1}(z_1) \chi_{1}(z_2)
 . . . \chi_{1}(z_{N_1}) \ \chi_{2}(w_1) \chi_{2}(w_2) . . . . . .\chi_{2}
 (w_{N_2}) \vert MF \rangle \eeqarr
In the above equation we have also inserted the matrix elements of 
 $R_{\alpha \beta}$ from eq (\ref{matrix}). 
 In terms of the composite fermion wavefunction defined in eq.(\ref{phifn})
we thus get
\beqarr \psi_{\nu_{1},\nu_{2}} (z_1 , z_2 ,. . . . . . . , z_{N_1} 
&;& \ w_1 , w_2 , . . . . . .  . , w_{N_2})  \nonumber \\
 &=& \prod_{i>j}^{N_1} (z_i - z_j)^{2s_{1}} \prod_{k>l}^{N_2}
  (w_k - w_l)^{2s_{2}} \prod_{r}^{N_1} \prod_{s}^{N_2} (z_r - w_s)^{n}
  \nonumber \\
&exp& \bigg[ {-1 \over 4l^2} \big(r_1 \sum_{1}^{N_1} 
    |z_i|^2 \ + \ r_2 \sum_{1}^{N_2} |w_i|^2) \bigg] \nonumber \\
&\phi&_{p_{1},p_{2}} (z_1 , z_2 ,. . . . . . . , z_{N_1} ;
 \ w_1 , w_2 , . . . . . .  . , w_{N_2}) \label {jaineqn2} \eeqarr
This is just the generalisation to double layers of Jain's formula relating
wavefunctions of electrons at certain   fractional fillings to corresponding
wavefunctions of composite fermions at other related fillings. 

\section{Discussion}

Several features of the result (\ref{jaineqn2}) are worth pointing out.

(a) Although this relation holds for any electronic filling
$\nu_{\alpha}$ and the corresponding composite fermion filling
$p_{\alpha}$, Jain's theory pertains to  cases  where the
$p_{\alpha}$ are integers.  Then, in the non-interacting limit,
the composite fermions will completely fill an integer number of
Landau levels, giving rise to an energy gap. Therefore the
usual arguments in the  quantum Hall literature can be invoked to
expect that even in the presence of e-e interactions, impurities
etc., an incompressible ground state will be obtained.

(b) The case $ \nu_2 \ = \ 0 $ cooresponds to no electrons at all in
the second layer, i.e. to the single layer case for which Jain proposed his 
ideas originally {\cite{Jain}}. For this case $N_2 \ = \ 0$ and the
 second-layer 
coordinates $w_i$ will be absent. Then (\ref{jaineqn2}), (\ref{reqn}) and 
(\ref{p-nu}) reduce to 
\beqarr \psi_{\nu_{1}} (z_1 , z_2 ,. . . . . . . , z_{N_1}) \  \  
 =\prod_{i>j}^{N_1} (z_i - z_j)^{2s_{1}} \ \  
exp \bigg[ {-1 \over 4l^2} \big(2s_{1} \nu_{1} \sum_{1}^{N_1} 
    |z_i|^2 \ \bigg] \nonumber \\
\phi_{p_1} (z_1 , z_2 ,. . . . . . . , z_{N_1} )
  \label {jaineqn} \eeqarr
with $\nu_1 = {p_1 \over (1+2p_{1}s_{1})} $
This is just Jain's well known formula for the single layer spinless problem.
That our procedure for constructing non-unitary transformations
to get flux-electron composites as
developed in \cite{Raj} will yield Jain's wavefunctions has also 
been pointed out by Wu and Yu \cite{Wu} for the single layer case.
Notice that the right hand side of eq.(\ref{jaineqn}) contains a
gaussian factor not included in Jain's version of this formula.
Whether such a factor should be there or not just depends on the
relative conditions under which the composite fermion system is
being compared to the electron system.  The way Jain writes such
an equation, the electron wavefunction $\psi$ and  the composite
fermion wave function $\phi$ correspond to the same magnetic
field. Hence there is no relative gaussian factor between them.
They also carry the same number of particles $N_{1}$. However they do 
correspond to different filling factors $p_{1}$ and $\nu_{1}$
and hence different densities from one another. This tacitly
implies that in Jain's way of writing this relationship the two
sides of the equation correspond to different areas (sample
sizes). By contrast , we have defined the electron operator
$\Psi_{\alpha}$ snd the composite fermion operator
$\chi_{\alpha}$ in the  same domain, as is natural in a field
theory. The sample areas  are thus taken as equal. The total number
of particles (and therefore the density) is also the same. The
difference in filling factors is caused by the difference in effective
magnetic fields, namely, B for the electron wavefunction $\psi$
and $B^*$ (as given in eq(\ref{B*})) for the composite fermion
state $\phi$. Hence the gaussian factors
(whose exponent is proportional to the magnetic field) will be different in
$\psi$ and $\phi$. The additional gaussian factor in (\ref{jaineqn})
 compensates for this difference. To verify this note that
  eq(\ref{B*}) reduces  for the single layer case, to
\beq \Delta B \ \equiv  \ B \ - \ B^* \  =  \   \ 
2s_{1} {\nu}_{1} B  \label{B1*} \eeq
Recalling that $B \propto  {1 \over l^2} $
we see that this difference $\Delta B $ corresponds precisely to the relative
 gaussian factor in (\ref{jaineqn}).

(c) Returning to the double layer system, the various filling fraction
possibilities contained in eq(\ref{nup-rel}) have been outlined at length
 by Lopez and Fradkin \cite{Lopez3} . Of particular interest is the case of 
$p_{1} \ = \ p_{2} \ = \ 1 $with $ s_{1}, s_{2} $ and n being arbitrary
integers.Then the filling fractions (\ref {nup-rel}) reduce to 
\beqarr \nu_{1} \ = \  {2s_2 + 1 - n \over (2s_1 + 1) \ (2s_2 + 1) - \ n^2 }
\nonumber \\ 
 \nu_{2} \ = \  {2s_1 + 1 - n \over (2s_1 + 1) \ (2s_2 + 1) - \ n^2 }
\label{nuhalp} \eeqarr
Our theory then yields for the corresponding electronic wavefunction
of this bilayer state with $\nu_{1}, \nu_{2}$ as given above,  the formula
(see \ref{jaineqn2}) :
\beqarr \psi_{\nu_{1},\nu_{2}} (z_1 , z_2 ,. . . . . . . , z_{N_1} 
; \ w_1 , w_2 , . . . . . .  . , w_{N_2})  \nonumber \\
 = \prod_{i>j}^{N_1} (z_i - z_j)^{2s_{1}} \prod_{k>l}^{N_2}
  (w_k - w_l)^{2s_{2}} \prod_{r}^{N_1} \prod_{s}^{N_2} (z_r - w_s)^{n}
 \  \ exp \bigg[ {-1 \over 4l^2} \big(r_1 \sum_{1}^{N_1} 
    |z_i|^2 \ + \ r_2 \sum_{1}^{N_2} |w_i|^2) \bigg] \nonumber \\
\phi_{1,1} (z_1 , z_2 ,. . . . . . . , z_{N_1} ;
 \ w_1 , w_2 , . . . . . .  . , w_{N_2}) \label {halp} \eeqarr
But $\phi_{1,1}$ is nothing but the wavefunction for unit
filling factor in each layer, for each of which we can use the 
$\nu = 1$ Laughlin wavefunction, corresponding to an effective magnetic
field of $B^{*}_{\alpha} =   \bigg( \ 1 \ - \   \ 
\sum_{\beta} R_{\alpha \beta} \nu_{\beta} \bigg) \ B
 \ = \ ( \ 1 - r_{\alpha}) B $. This Laughlin wavefunction for unit filling
in the first layer is
\beq \psi^{\nu = 1}_{Laughlin} \ = \ \prod_{i>j}^{N} (z_i - z_j)
exp \bigg[ {-1 \over 4l_{1}^2} \big( \sum_{1}^{N} |z_i|^2 \bigg] \eeq
where 
\beq {1 \over l_{1}^2} \ \equiv \ {eB^{*}_{1} \over \hbar  c}
 \ = \ {(1 - r_{1}) \over l^2} \eeq
 and similarly for the
second layer in terms of the coordinates $w_i$ . Inserting such a 
$\phi_{1,1}$ into (\ref{halp}) we get
 \beqarr \psi_{\nu_{1},\nu_{2}} (z_1 , z_2 ,. . . . . . . , z_{N_1} 
&;& \ w_1 , w_2 , . . . . . .  . , w_{N_2})  \nonumber \\
 &=& \prod_{i>j}^{N_1} (z_i - z_j)^{2s_{1}+1} \prod_{k>l}^{N_2}
  (w_k - w_l)^{2s_{2}+1} \prod_{r}^{N_1} \prod_{s}^{N_2} (z_r - w_s)^{n}
  \nonumber \\
&exp& \bigg[ {-1 \over 4l^2} \big( \sum_{1}^{N_1} 
    |z_i|^2 \ + \  \sum_{1}^{N_2} |w_i|^2) \bigg] \label {halp1} \eeqarr

This is just the Halperin wavefunction for the ground state of
the bilayer system \cite{GirvMac} , \cite{Halp} with filling
factors in the two layers given by (\ref{nuhalp}). Such a
wavefunction was derived from a Chern-Simons field theoretic
model by Ezawa and Iwasaki \cite{Ezawa} earlier, but in a very
different way. Their resuts are based on solutions to certain
semiclassical self dual equations which require not only a
mean-field approximation but also the neglect of e-e Coulomb
interactions. They do treat interactions , but perturbativey, in the short
distance limit. Our work here does use a mean field 
approximation in deriving results such as (\ref{halp}), but nowhere
has the Coulomb term in the Hamiltonian dropped nor short distance
 appromation made. We also provide
exact operator definitions of the composite particles.  This wavefunction 
 for the  case $2s_1 + 1 = 2s_2 + 2
$ (which is a special case of the above ) was also obtained by
Lopez and Fradkin \cite{Lopez3}. They used  their fermionic Chern-Simons
 theory \cite{Lopez1} suitably
generalised to two-component wavefunctions and obtained the modulus of
the wavefunction in the long wavelength limit. Our
field theory, using the composite operators defined in
(\ref{chipi})  gives the full wavefunction including its
modulus, phase and gaussian factors. No long wavelength approximation has
 been invoked by us.

(d) Clearly our work is only a variant on the earlier work of Zhang et al, 
Read and   Lopez and Fradkin cited above. The main advantage of our 
generalisation to non-unitary transformations and complex Chern Simons fields
is that our composite operators incorporate all aspects of the Laughlin and 
Jain wavefunctions including the moduli $| z_i - z_j | $ of the 
correlations. These moduli contain the important zeroes which should be there 
in the presence of Coulomb repulsion, and also restore the wavefunctions to
the lowest Landau level. The gaussian factors which should be present in the 
wavefunctions are also incorporated. Indeed, as our derivation shows these
gaussian factors are essential for cancelling the imaginary part of our 
complex statistical gauge field in the mean field limit.
True, some of these factors are
 obtained in the
works of Zhang et al \cite {ZHK} and Lopez and Fradkin \cite {Lopez3}
but only upon including fluctuations about the mean field. That they
are present already at the mean field level in our operators indicates
that our operators may be better candidates for flux-electron composite
fields.

(e) That we already get the full Laughlin and Jain wavefunctions at the
 mean field level raises hopes that corrections to these wavefunctions could be
 obtained even at the lowest order in fluctuations about the mean field.
 Unfortunately, this is where the non-unitary nature of our transformation
 (\ref{chipi}) could create difficulties. Although the imaginary parts
 of our statistical field $\vec v({\vec x})$ cancel in the mean field
  approximation they will be present away from mean field. Of course the full 
  Hamiltonian (\ref{HB}) is hermitian as can be verified using
   (\ref{chipirelation}),
   but its separation into a mean field part and a perturbation 
 does not maintain hermiticity in each part.   
  Standard perturbation techniques would have to be re-examined and modified
 to take this into account. These remarks hold not only for the
 present work but also our earlier work with Sondhi \cite{Raj}. For
  a discussion on how to go beyond mean field theory in such cases see
  Wu and Yu \cite {Wu}.

\end{document}